# The Vitruvius' Tale of Archimedes and the Golden Crown

Amelia Carolina Sparavigna
Dipartimento di Fisica,
Politecnico di Torino, Torino, Italy

The paper discusses the tale that we can find in "The Architecture" by Vitruvius, on a method used by Archimedes to determine the percentage of gold and silver in a crown. The method is based on the immersion of bodies, allowing the evaluation of their volume in the case of irregular shapes. The measurement, as reported in "The Architecture", seems to be difficult to realize. But, using a vessel for a water-clock, the approach that Vitruvius described is possible. Here the discussion and experiments.

Marcus Vitruvius Pollio, who lived during the first century BC, was a Roman writer, architect and engineer. He wrote a book on architecture, "The Architecture", that he dedicated to the emperor Augustus [1]. Surviving from the classical antiquity, this book inspired several architects and artists of the Renaissance. "The Architecture" is also the source of one of the most famous and popular anecdote on Archimedes of Syracuse, one of the greatest scientists in classical antiquity. Greek mathematician and physicist, inventor and astronomer, Archimedes lived in the third century BC, and died during the Siege of Syracuse, killed by a Roman soldier.

Vitruvius tells us that Archimedes devised a method for measuring the volume of objects with irregular shape: this description is also known as the tale on the golden crown. According to Vitruvius, King Hiero of Syracuse ordered a votive crown for a temple. Having some doubts, the king asked Archimedes to determine whether some silver had been used by the goldsmith. Since the object was quite beautiful, he asked to avoid damaging the crown. The scientist was quite troubled, because he had to determine the volume of this irregular object without melting it. But, when he was taking a bath, noticing that the level of the water increased when he entered the vessel, he immediately realized that this effect could be used to measure the volumes. Excited by the discovery, he cried out "Eureka!, that is "I have found it!". The Vitruvius' tale continues describing the method of measurement.

As reported in Ref.[2], this "story of the golden crown does not appear in the known works of Archimedes". The item [2] continues telling that "the practicality of the method it describes has been called into question, due to the extreme accuracy with which one would have to measure the water displacement" [3]. According to [2], Archimedes may have applied his knowledge of hydrostatics, the Archimedes' Principle that he described in his book On Floating Bodies, to fulfil the Hiero's request. Using this principle, he could have compared the density of the golden crown to that of gold by using a scale as shown in [2], in a procedure that Galileo considered as probably "the same that Archimedes followed, since, besides being very accurate, it is based on demonstrations found by Archimedes himself" [2,3].

The method described by Vitruvius is then questioned. In any case, I asked myself: is it possible that Vitruvius described a method so different from that used by Archimedes? In any case, is the Vitruvius' description useful for measurement?

First of all, let us see what Vitruvius is writing in the third chapter, "of the method of detecting silver when mixed with gold" [1]. "Charged with this commission (to determine whether the crown had silver inside or not), he (Archimedes) by chance went to a bath, and being in the vessel, perceived that, as his body became immersed, the water ran out of the vessel. Whence, catching at the method to be adopted for the solution of the proposition, he immediately followed it up, leapt out of the vessel in joy, and, returning home naked, cried out with a loud voice that he had found that of which he was in search, for he continued exclaiming, in Greek, Eureka, (I have found it out). After this, he is said to have taken two masses, each of a weight equal to that of the crown, one of them of gold and the other of silver. Having prepared them, he filled a large vase with water up to the brim, wherein he placed the mass of silver, which caused as much water to run out as was equal to the bulk thereof. The mass being then taken out, he poured in by measure as much water as was required to fill the vase once more to the brim. By these means he found what quantity of water was equal to a certain weight of silver. He then placed the mass of gold in the vessel, and, on taking it out, found that the water which ran over was lessened, because, as the magnitude of the gold mass was smaller than that containing the same weight of silver. After again filling the vase by measure, he put the crown itself in, and discovered that more water ran over then than with the mass of gold that was equal to it in weight ; and thus, from the superfluous quantity of water carried over the brim by the immersion of the crown, more than that displaced by the mass, he found, by calculation, the quantity of silver mixed with the gold, and made manifest the fraud of the manufacturer."

Let us see now the calculations in Ref.3, which is questioning the Vitruvius' description. This reference assumes the method described by Vitruvius is based on a simple immersion of bodies. It is supposed that the Hiero's crown weighed 1000 grams. Gold has a density of 19.3 grams/cm$^3$, 1000 grams of gold would have a volume of 51.8 cm$^3$. As an example, the reference continues considering that the goldsmith replaced a 30% (300 grams) of the gold in the crown by silver. The density of silver is of 10.5 grams/cm$^3$. As a consequence, the gold-silver crown has a volume of 64.8 cm$^3$. In fact, in this example, the difference of volumes is more than 10 cm$^3$.

Ref.3 supposes a vessel used in the experiment, with a circular opening with a diameter of 20 cm. The opening has then a cross-sectional area of 314 cm$^2$. The solid gold immersed in the vessel would raise the level of water at the opening by 0.165 cm. The crown instead would raise the level of the water at the opening by 0.206 cm. The difference in the level of water, displaced by the crown and the solid gold is of 0.41 mm. Of course, this difference is too small to be directly observed. Other sources of error are the surface tension, the water remaining on the objects after removal and air bubbles. After this discussion, this method of measurement is discharged.

In fact a simple immersion in a vessel is not suitable for measurements. But Vitruvius is not describing a simple immersion method, but a method based on the overflow of water at the rim of the vessel (labra) and replacement of water in the vessel, evaluated by means of some small measuring vessels (sextario mensus, for the Latin text, see Ref.4). If we imagine a classic marble vessel with a smooth rim, the overflow of the water is difficult to control. But Archimedes had surely a more suitable device, to measure the water displaced by immersion. In my opinion, it is the vessel used for the water-clock (see Fig.1), that is, a vessel with a hole near the rim. Let us remember that a water-clock, or clepsydra, is a device in which time is measured by the flow of water. Water clocks, along with sundials, are assumed to be the oldest time-measuring devices, "with the only exceptions being the vertical gnomon and the day-counting tally stick" [5]. For sure, the simplest form of water-clocks existed in Babylon and Egypt, around the 16th century BC. The Greeks improved the water clock design.

Let us assume that Archimedes used a vessel with a hole. I have tried to repeat what is described by Vitruvius in the following manner. I used two transparent plastic cups, one with a hole, with a plastic ring glued about it. It simulates the vessel of a water-clock. Then I used a plumb (400 grams) and a weight of a balance of 100 grams for immersions.

The initial levels of water in the two cups is shown in Fig.3A. The level in the right cup is determined by its hole. The level in the left cup is marked on it by a pen. The plumb (red) is inserted in the cup with the hole; with a slow immersion of the object, the water flow out through the hole, till its level is again that determined by the hole (Fig.3B). The red plumb is carefully removed from the cup, avoiding that some water remains on it. The level in the cup is lower (see Fig.3C). By means of a little spoon, water is passed from the left cup to that with the hole, until a drop starts to pass through the hole (Fig.3D). To check that the water transferred from left to right is almost equal, the plumb-bob is inserted in the left cup. We see that the water is again at the original marked level (see Fig3E,F). Note that the volume of the plumb-bob is comparable to that of 1000 grams of gold.

I have repeated the procedure done for the red plumb with a smaller object, the weight of a balance (100 grams). It is approximately 10 cm$^3$ in volume, that is, equal to the supposed difference of volumes between the crown and the solid gold. The initial levels of water in the two cups is shown in panel A of Fig.4. The weight (bronze) is inserted in the cup with the hole; again, the water flows out through the hole, till its level is that determined by the hole (Fig.4B). The mass is carefully removed from the cup, avoiding that some water remains on it. The level in the cup is lower. As before, using the little spoon, some water is moved from the left cup to the right, until a drop starts flowing through the hole. The final levels are shown in Fig.4C. To check that the water transferred from left to right is almost equal, the weight is inserted in the left cup. We see that the level of water is again at the original marked one (Fig.4D).

Using a stick with several cuts, it is easy to measure the level in the left cup and then evaluate the volume of water removed from it, to replace the water lost by the vessel with the hole after immersion of bodies. This is the method described by Vitruvius, if we consider a vessel for water-clocks as that used in the measurements. As Fig.3 and 4 are showing, in particular Fig.4, it seems that Archimedes could have easily estimated the difference of volumes of the crown and solid gold, in the manner that Vitruvius is reporting, if this difference was of about 10 cm$^3$. In fact, as shown in Fig 4, a volume of about 10 cm$^3$ can be easily appreciated, but also smaller volumes of a few cubic-centimetres. Of course, there are some sources of errors as told in Ref.3, but they can be reasonably reduced.

Probably, the method described by Vitruvius was the first that Archimedes used in his measurements on the densities of bodes, to give a rapid answer to his king's request. In the following, as Galileo considered, he could have devised more precise methods involving a balance.

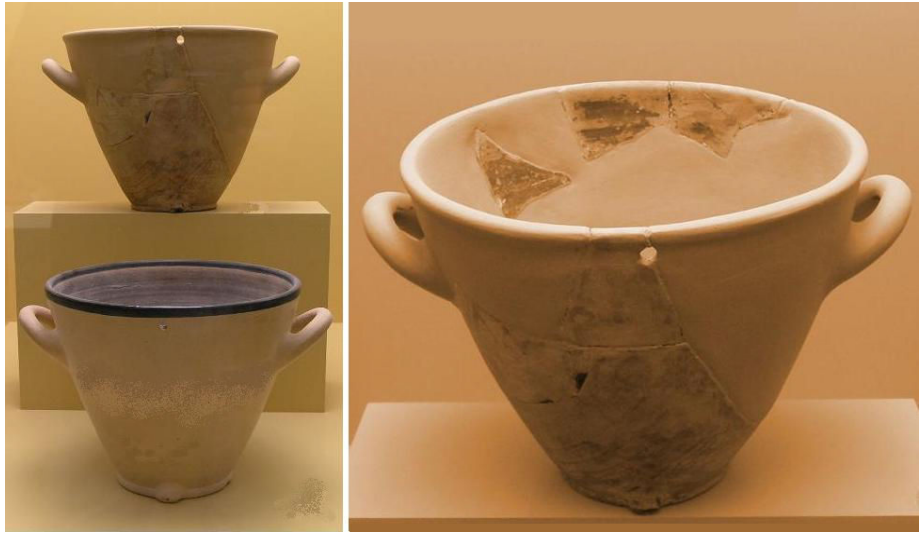

Fig.1 Archimedes had surely some vessels as those shown in the figure. They are those used for water clocks. Note the hole near the rim. A water clock, or clepsydra, is a device in which time is measured by a regulated flow of water. Images are adapted from those of courtesy at Wikipedia.

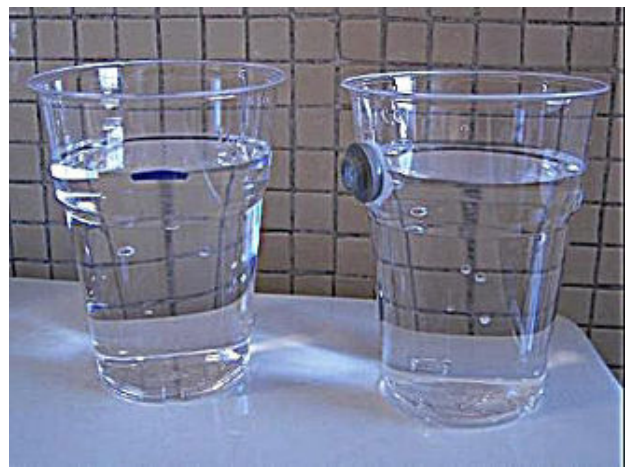

Fig.2 Two cups can be used to repeat the measure of volumes as described by Vitruvius. On the right, we see the cup with a hole, with a plastic ring glued about it. It simulates the vessel of a water-clock

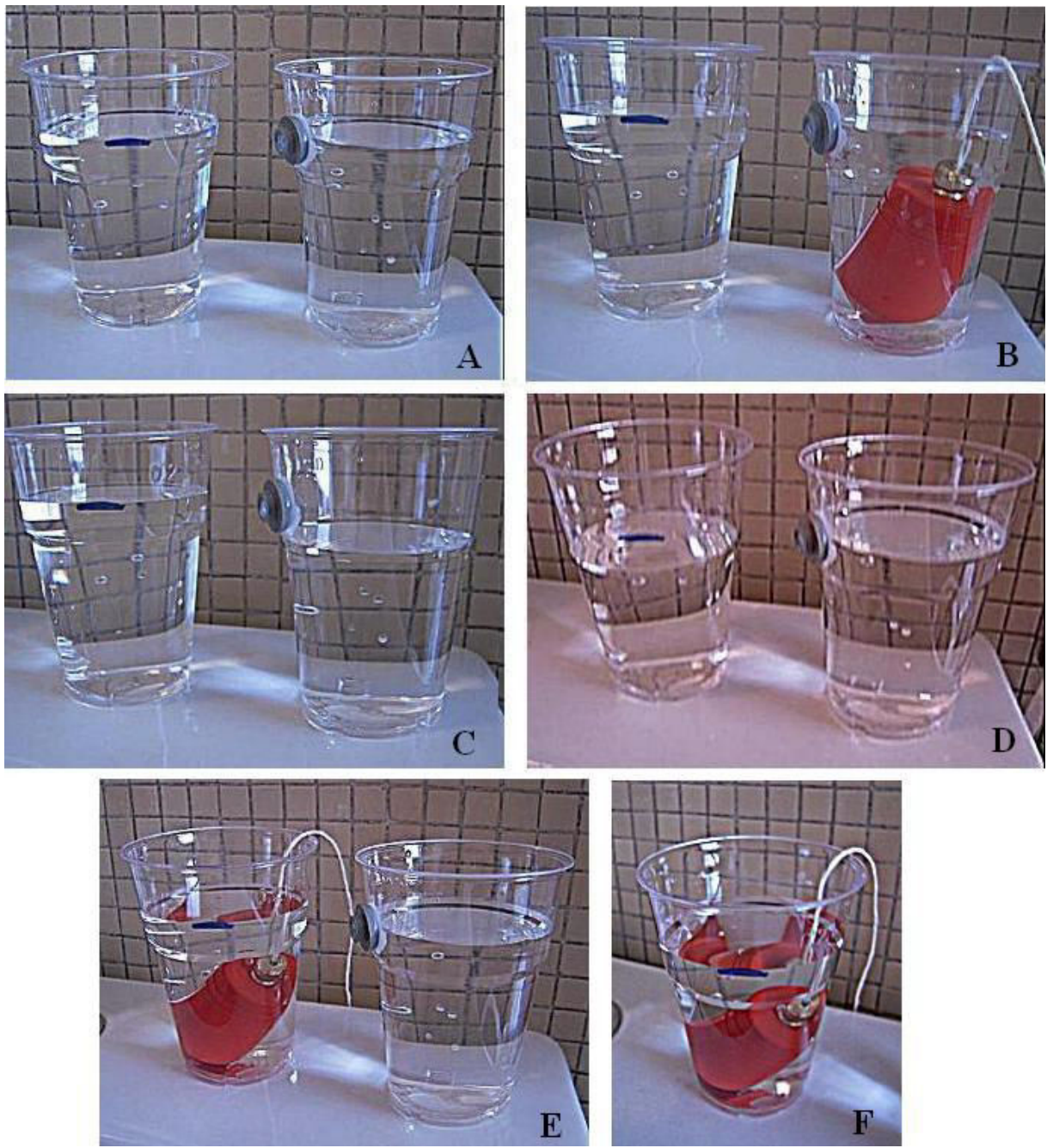

Fig.3. The initial levels of water in the two cups is shown in panel A. The level in the right cup is determined by the hole. The level in the left cup is marked on it by a pen. The plumb (red) is slowly inserted in the cup with the hole; the water flow out through the hole, until its level is again that determined by the hole (B). The red plumb is carefully removed from the cup, avoiding that some water remains on it. The level in the cup is lower (C). By means of a little spoon, water is passed from the left cup to the right one, until a drop flows through its hole (D). To check that the water transferred from left to right is almost equal, the plumb-bob is inserted in the left cup. We see that the level of water is again at the original marked level (E,F).

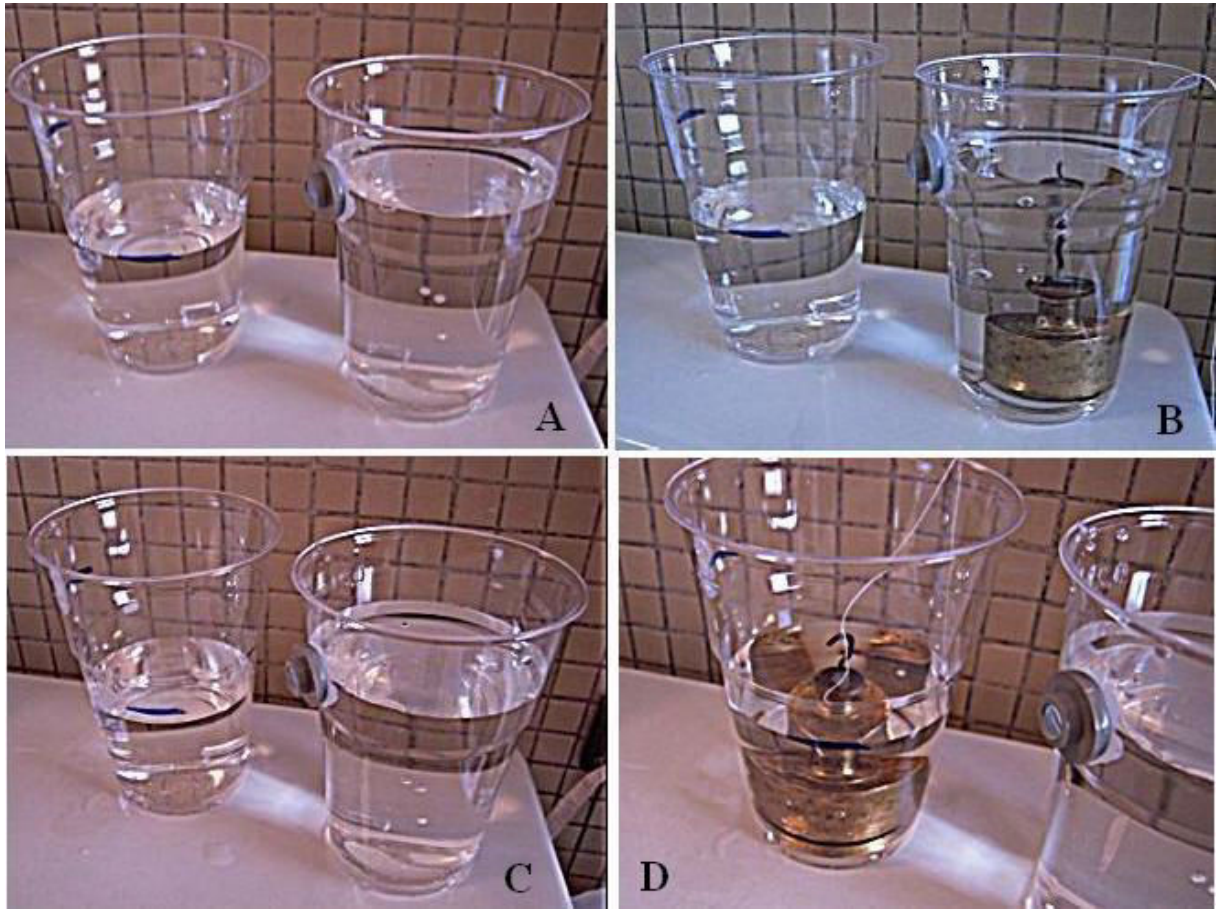

Fig.4. The procedure shown in Fig.3 is repeated with the weight of a balance (100 grams). The initial levels of water in the two cups is shown in panel A. The weight (bronze) is inserted in the cup with the hole; the water flows out through the hole, until its level is again that determined by the hole (B). The mass is carefully removed from the cup, avoiding that some water remains on it. The level in the right cup is lower. By means of a little spoon, water is passed from the left cup to that with the hole, until a drop flows through it (C). To check that the water transferred from left to right is almost equal, the weight is inserted in the left cup. We see that the level of water is again at the original marked level (D).